\documentclass[aps,prd,twoside,onecolumn,superscriptaddress,floatfix,float,nofootinbib,showpacs]{revtex4}
\usepackage{amsmath}
\usepackage{bm}
\usepackage{xcolor}
\usepackage{hyperref}
\usepackage{url}
\usepackage{graphicx}
\usepackage[large]{subfigure}
\usepackage{amssymb}
\usepackage[amssymb]{SIunits}
\usepackage{aas_macros}

\usepackage{multirow}

\newcommand\beqn{\begin{eqnarray}}
\newcommand\eeqn{\end{eqnarray}}

\newcommand\lsim{\mathrel{\rlap{\lower4pt\hbox{\hskip1pt$\sim$}}
        \raise1pt\hbox{$<$}}}
\newcommand\gsim{\mathrel{\rlap{\lower4pt\hbox{\hskip1pt$\sim$}}
        \raise1pt\hbox{$>$}}}

\usepackage[mathscr]{eucal}	
\DeclareTextSymbol{\degre}{T1}{23}

\newcommand{\vl} { {\mbf{\ell}} }
\newcommand{\vL} { {\mbf{L}} }

\newcommand{\beq} {\begin{equation}}
\newcommand{\eeq} {\end{equation}}
\newcommand{\bal} {\begin{aligned}}
\newcommand{\eal} {\end{aligned}}

\newcommand{\mbf}[1]{\mbox{\boldmath$#1$}}

\begin{document}

\title{Bias to CMB lensing from lensed foregrounds}

\author{Nishant Mishra}
\email{nishant.mishra@berkeley.edu}
\affiliation{Department of Physics, University of California, Berkeley, CA 94720, USA}
\author{Emmanuel Schaan}
\email{eschaan@lbl.gov}
\affiliation{Lawrence Berkeley National Laboratory, One Cyclotron Road, Berkeley, CA 94720, USA}
\affiliation{Berkeley Center for Cosmological Physics, University of California, Berkeley, CA 94720, USA}

\begin{abstract} 

Extragalactic foregrounds 
are known to constitute a limiting systematic in temperature-based CMB lensing with AdvACT, SPT-3G, Simons Observatory and CMB S4.
Furthermore, since these foregrounds are emitted at cosmological distances, they are also themselves lensed.
The correlation between this foreground lensing and CMB lensing 
causes an additional bias in CMB lensing estimators.
In this paper, we quantify for the first time this ``lensed foreground bias'' for the standard CMB lensing quadratic estimator, the CMB shear and the CMB magnification estimators,
in the case of Simons Observatory
and in the absence of multi-frequency component separation.
This percent-level bias is highly significant in cross-correlation of CMB lensing with LSST galaxies,
and comparable to the statistical 
uncertainty in CMB lensing auto-spectrum.
We discuss various mitigation strategies, and show that ``lensed foreground bias-hardening'' methods can reduce this bias at some cost in signal-to-noise.
The code used to generate our theory curves is publicly available\footnote{\url{https://github.com/EmmanuelSchaan/LensedForegroundBias}}.
\end{abstract}

\maketitle

\section{Introduction}

Gravitational lensing of the cosmic microwave background (CMB) probes the projected mass distribution in the Universe, all the way to the surface of last scattering.
This effect is measured at high significance from the WMAP satellite \cite{2007PhRvD..76d3510S, 2008PhRvD..78d3520H}, the Atacama Cosmology Telescope (ACT) \cite{2011PhRvL.107b1301D, 2014JCAP...04..014D, 2015PhRvL.114o1302M, 2015ApJ...808....7V, 2017PhRvD..95l3529S, 2015MNRAS.451..849A}, the South Pole Telescope (SPT) \cite{2012ApJ...756..142V, 2013PhRvL.111n1301H, 2015ApJ...806..247B, 2015ApJ...810...50S, 2015ApJ...806..247B, 2017ApJ...849..124O}, POLARBEAR \cite{2014PhRvL.113b1301A, 2019arXiv190402116N, 2019arXiv190307046A}, the Planck satellite \cite{2014A&A...571A..17P, 2015arXiv150201591P, 2018arXiv180706210P} and BICEP 2 / Keck Array \cite{2016arXiv160601968K}.
These measurements contain valuable information on the nature of dark energy and the masses of the neutrinos, through their effect on the growth of structure and the expansion history in the universe.
Via delensing \cite{2017ApJ...846...45M, 2017JCAP...05..035C, 2019PhRvD..99d3518C} of the CMB polarization B-modes, these lensing measurements will also be crucial in the search for primordial gravitational waves.

As the statistical signal-to-noise (SNR) in upcoming CMB lensing detections increases with AdvACT \cite{2016JLTP..184..772H}, SPT-3G \cite{2014SPIE.9153E..1PB}, Polarbear-2 and Simons Array \cite{2016JLTP..184..805S}, Simons Observatory \cite{2019JCAP...02..056A} and
CMB S4 \cite{2016arXiv161002743A},
a similar improvement in systematics control becomes necessary.
In temperature-based CMB lensing reconstruction,
extragalactic foregrounds such as the cosmic infrared background (CIB), the thermal and kinematic Sunyaev-Zel'dovich (tSZ and kSZ) effects and radio point sources (radio PS) constitute the main limiting systematics.
If not accounted for, they are known to produce highly significant biases to CMB lensing \cite{2014JCAP...03..024O, 2014ApJ...786...13V, 2017JCAP...08..030R, 2018PhRvD..97b3512F, 2019PhRvL.122r1301S}.
These biases arise from the non-Gaussian statistics of these foregrounds, and their correlation with CMB lensing.
Since these foregrounds dominate on small scales, they limit the range of temperature multipoles that can be used for lensing reconstruction to $\ell < 3000-3500$.
Various mitigation methods exist.
For quadratic CMB lensing estimators, the lensing field is reconstructed from two powers of the temperature map, one ``gradient leg'' and one small-scale leg. 
Mitigation methods typically aim at removing the foreground from one or two of these powers of the map.
Masking removes the brightest point sources in the temperature map, and is an effective way to control radio PS \cite{2014ApJ...786...13V}.
Multi-frequency component separation \cite{2019ApJ...872..170R} or the cleaned gradient method \cite{2018arXiv180208230M} can reduce the amplitude of CIB and tSZ at some cost in noise, but cannot reduce the kSZ bias.
In the case of halo lensing, the gradient can also be foreground-cleaned by inpainting a Gaussian CMB realization at the localization of the halo \cite{2019arXiv190413392R}.
In some cases, the foregrounds are left intact in the two legs of the quadratic estimator, and the mitigation occurs by nulling the overall response of the estimator to the foreground.
``Bias-hardened'' estimators reduce the contamination from foregrounds whose statistics is known (e.g., Poisson) \cite{2013MNRAS.431..609N, 2014MNRAS.438.1507N, 2014JCAP...03..024O}.
Splitting the CMB lensing quadratic estimator into magnification-only and shear-only estimators \cite{2012PhRvD..85d3016B, 2018JCAP...01..034P, 2019PhRvL.122r1301S} provides a useful foreground null test.
Discarding the magnification-only part appears to significantly reduce the contamination from all extragalactic foregrounds, at a cost in SNR which can be compensated by including smaller temperature multipoles in the shear estimator \cite{2019PhRvL.122r1301S}.
The ``multipole estimators'', which generalize this decomposition, are expected to share the same property \cite{2019PhRvL.122r1301S}.
These various methods are based on different principles. 
Multi-frequency foreground cleaning and the cleaned gradient method partially remove the foreground from the temperature map (or one of the two, for the cleaned gradient method), so that the foregrounds do not enter the lensing estimator or do not bias it.
Masking works because a large part of the non-Gaussianity of some foregrounds (radio PS, CIB, tSZ) comes from localized peaks associated with individual galaxies or halos. 
Masking or inpainting these foreground peaks does not completely remove the foreground from the temperature map, but it reduces its non-Gaussianity (bispectrum and trispectrum).
With bias hardening and the shear/multipole estimators, the foreground is left intact in the temperature map and fed to the lensing estimator. However, the estimator is modified so as to be insensitive to the foreground bispectrum and trispectrum.
The best solution to the lensing biases from the non-Gaussianity of extragalactic foregrounds likely involves combining these different methods.

However, extragalactic foregrounds produce another bias to CMB lensing, discussed in \cite{2004NewA....9..173C} but not yet been quantified, to the best of our knowledge.
This bias is not a consequence of the non-Gaussianity of the foregrounds, but of the fact that extragalactic foregrounds are themselves lensed, by an amount correlated with CMB lensing \cite{2004NewA....9..173C, 2018PhRvD..97l3539S, 2019arXiv190502084F}.
Indeed, extragalactic foregrounds are emitted at cosmological distances, and should therefore be distorted by lensing, just like the CMB and galaxy shapes are.
CMB lensing quadratic estimators will generically reconstruct not only the lensing of the CMB, but also the lensing of the foregrounds \cite{2018PhRvD..97l3539S}. 
The correlation of CMB lensing and foreground lensing enhances this bias to CMB lensing.
This bias is present even if the extragalactic foregrounds are perfectly Gaussian, as long as they are lensed by an amount correlated with CMB lensing.
In this paper, we quantify this bias for the first time, and show that it is significant for an experiment like Simons Observatory.
This bias will be even larger for the CMB S4 temperature-based lensing. 
However, the polarization-based quadratic estimators will carry a larger weight for CMB S4, and extragalactic foregrounds are expected to be less important there.

\section{Review: CMB lensing reconstruction with quadratic estimators}

\subsection{CMB lensing convergence}

We consider the CMB to be emitted at a single redshift at the last scattering surface.
For a source image at a single redshift and in the Born approximation, the lensing convergence is obtained as
\beq
\kappa^\text{CMB}_{\vL} =  
\int_0^{\chi_\text{CMB}} d\chi \; 
W^{\kappa} (\chi, \chi_\text{CMB}) \; \delta_m (\vec{k} = \vL/\chi, \chi),
\eeq
where $W^\kappa(\chi, \chi_S)$ is the lensing kernel for a source image at distance $\chi_S$:
\beq
W^\kappa (\chi, \chi_S) = \frac{3}{2} \left( \frac{H_0}{c} \right)^2 \Omega_m^0 \frac{\chi}{a(\chi)} \left( 1 - \chi/\chi_S \right).
\eeq
In the Limber and flat sky approximations, the CMB lensing power spectrum thus becomes:
\beq
C_L^{\kappa_\text{CMB}}
=
\int \frac{d\chi}{\chi^2} \;
W^{\kappa}(\chi, \chi_\text{CMB})^2
P_m \left(k = \frac{L+1/2}{\chi}, z(\chi)\right),
\eeq
where $P_m$ is the nonlinear matter power spectrum, computed using the \texttt{Halofit} \cite{2015MNRAS.454.1958M} implementation in \texttt{CLASS} \cite{2011JCAP...07..034B}.

\subsection{Quadratic CMB lensing estimators}

Throughout this paper, we adopt the flat sky approximation, and decompose the various maps (CMB, foreground, convergence) in Fourier modes rather than spherical harmonics. We denote by $T^0$ an unlensed map (CMB or foreground), and $T$ the corresponding lensed map.
Lensing produces off-diagonal correlations in the observed temperature map:
\beq
\langle T_\vl T_{\vL-\vl} \rangle
=
f_{\vl, \vL-\vl} \kappa_\vL
+
\mathcal{O}\left( \kappa^2 \right),
\label{eq:lensed_tt}
\eeq
where the response function $f_{\vl, \vL-\vl}$ is completely determined by the unlensed power spectrum:
\beq
f_{\vl_1, \vl_2}
\equiv
\left(
\frac{2 \vL}{L^2}
\right)
\cdot
\left[
\vl_1  C^0_{\ell_1}
+
\vl_2  C^0_{\ell_2}
\right].
\eeq
In particular, we shall call this response $f^\text{CMB}$ when $C^0$ is the unlensed CMB power spectrum, $f^f$ when $C^0$ is the unlensed foreground power spectrum, and $f^{ff'}$ when $C^0$ is the cross-spectrum of two correlated unlensed foregrounds, such as CIB and tSZ.

This coupling of Fourier modes is used to construct unbiased CMB lensing quadratic estimators as
\begin{equation}
\mathcal{Q}_\vL \left[ T, T \right]
=
N_\vL
\int \frac{d^2 \vl}{(2\pi)^2} T_\vl T_{\vL-\vl} F_{\vl, \vL-\vl},
\label{eq:quadratic_estimator}
\end{equation}
where the weight function $F_{\vl, \vL-\vl}$ can in principle be chosen arbitrarily,
and the corresponding normalization is then fixed in order to obtain unit response to CMB lensing:
\beq
N_\vL = \left(\int \frac{d^2 \vl}{(2\pi)^2} F_{\vl, \vL-\vl} f^\text{CMB}_{\vl, \vL-\vl} \right)^{-1}.
\label{eq:normalization}
\eeq
In particular, the response function $f^\text{CMB}_{\vl, \vL-\vl}$ is that of the CMB,
so that the estimator has unit response to lensing when applied to lensed CMB maps.
In practice, the weights $F_{\vl, \vL-\vl}$ are typically chosen so as to minimize the variance of the estimator \cite{2002ApJ...574..566H}.
These weights can be chosen differently, for example to null the response to point sources or the survey mask as in ``bias hardening'' \cite{2013MNRAS.431..609N, 2014MNRAS.438.1507N, 2015arXiv150201591P, 2018JCAP...07..046F, 2018arXiv180706210P}, to discard the information in shear or magnification \cite{2008MNRAS.388.1819L, 1999PhRvD..59l3507Z, 2004NewA....9..417P, 2010PhRvD..81l3015L, 2012PhRvD..85d3016B, 2018JCAP...01..034P, 2019PhRvL.122r1301S}, or to discard contaminated modes in one of the legs of the estimator \cite{2018arXiv180208230M}.
In this paper, we will consider the following quadratic estimators:
\beq
F_{\vl, \vL-\vl}
=
\left\{
\bal
&\frac{f^\text{CMB}_{\vl, \vL-\vl}}{2 C^\text{total}_\ell C^\text{total}_{|\vL-\vl|}}
&&\text{QE}\\
&\frac{C^\text{CMB}_\ell}{2(C^{\text{total}}_\ell)^2} \; \frac{d \ln C^\text{CMB}_\ell}{d \ln \ell} \cos(2\theta_{\vL, \vl})
&&\text{Shear}\\
&\frac{C^\text{CMB}_\ell}{2(C^{\text{total}}_\ell)^2} \; \frac{d \ln \ell^2C^\text{CMB}_\ell}{d \ln \ell}
&&\text{Magnification}\\
\eal
\right.
,
\eeq
where $C^\text{CMB}$ is the unlensed CMB power spectrum and $C^{\text{total}}$ is the total map power spectrum, including lensed CMB, foregrounds and detector noise.
In what follows, for convenience, we further symmetrize these lensing weights with the substitution 
$F_{\vl, \vL-\vl} \rightarrow \left( F_{\vl, \vL-\vl} + F_{\vL-\vl, \vl} \right) / 2$.

As shown in Fig.~2 of \cite{2019PhRvL.122r1301S} and explained in \cite{2019PhRvL.122r1301S}, the noise power spectrum of the shear and magnification estimators shows a spike at $L\sim 3000$. 
In all the calculations below, the error bars indeed display this spike at that multipole, as expected.
This is due to the lensing weights $F_{\ell, L-\ell}$ for the shear and magnification only being optimal in the large-scale lens regime $L \ll \ell$, where large-scale lensing modes are reconstructed from small-scale temperature modes. Outside of this regime, the lensing weights $F_{\ell, L-\ell}$ are effectively arbitrary, and lead to a null response to lensing at $L\sim 3000$, causing the spike in the noise power spectrum.

\section{Extragalactic foregrounds and their lensing}

Intuitively, the lensed foreground bias should depend on several properties of the foreground.
First, the size of the bias depends on the amplitude of the foreground power spectrum.
For example, if the foreground power spectrum is multiplied or divided by some factor, the bias to the CMB lensing estimators is also multiplied or divided by the same factor, since these estimators are quadratic in the temperature. 
Second, the bias also depends on the shape of the foreground power spectrum. 
For instance, if a foreground component had the same exact power spectrum as the CMB, then the CMB lensing estimators would have unit response not only to CMB lensing, but also to the foreground lensing.
The shape of the foreground power spectrum thus determines how much the CMB lensing estimators respond the foreground lensing.
Finally, the redshift distribution of the foreground sources determines the amplitude of the foreground lensing convergence, and the size of its correlation with the CMB lensing convergence. 
In this section, we estimate each of these foreground properties.

\subsection{Experimental configuration \& foreground power}

Throughout the paper, we consider a ``CMB S3'' experiment, similar to Simons Observatory, with a beam full-width at half maximum of $1.4'$ and a white noise level of $7 \mu K'$ in temperature at 143 GHz.
The lensing reconstruction for the QE, shear and magnification estimators uses all Fourier modes with $\ell_\text{min, T} = 30 \leq \ell \leq \ell_\text{max, T} = 3,500$.
We consider a temperature map at a single frequency, $143$ GHz, without multi-frequency component separation.
This slightly pessimistic assumption will produce slightly larger foreground biases, but it makes our results independent of the particular choice of component separation method.
Furthermore, the shear estimator was shown to be robust to foreground contamination \cite{2019PhRvL.122r1301S}, so it may be used on a single-frequency temperature map.
In the lensing weights defined above, the total power spectrum thus includes not only the lensed CMB and the white detector noise, but also all the foreground power spectra.
This realistic choice downweights the small scales where foregrounds dominate, and will thus reduce the foreground bias.
We do not include noise from atmospheric emission, as this quantity varies with observing site.

\begin{figure}[h!!!!]
\includegraphics[width=0.45\textwidth]{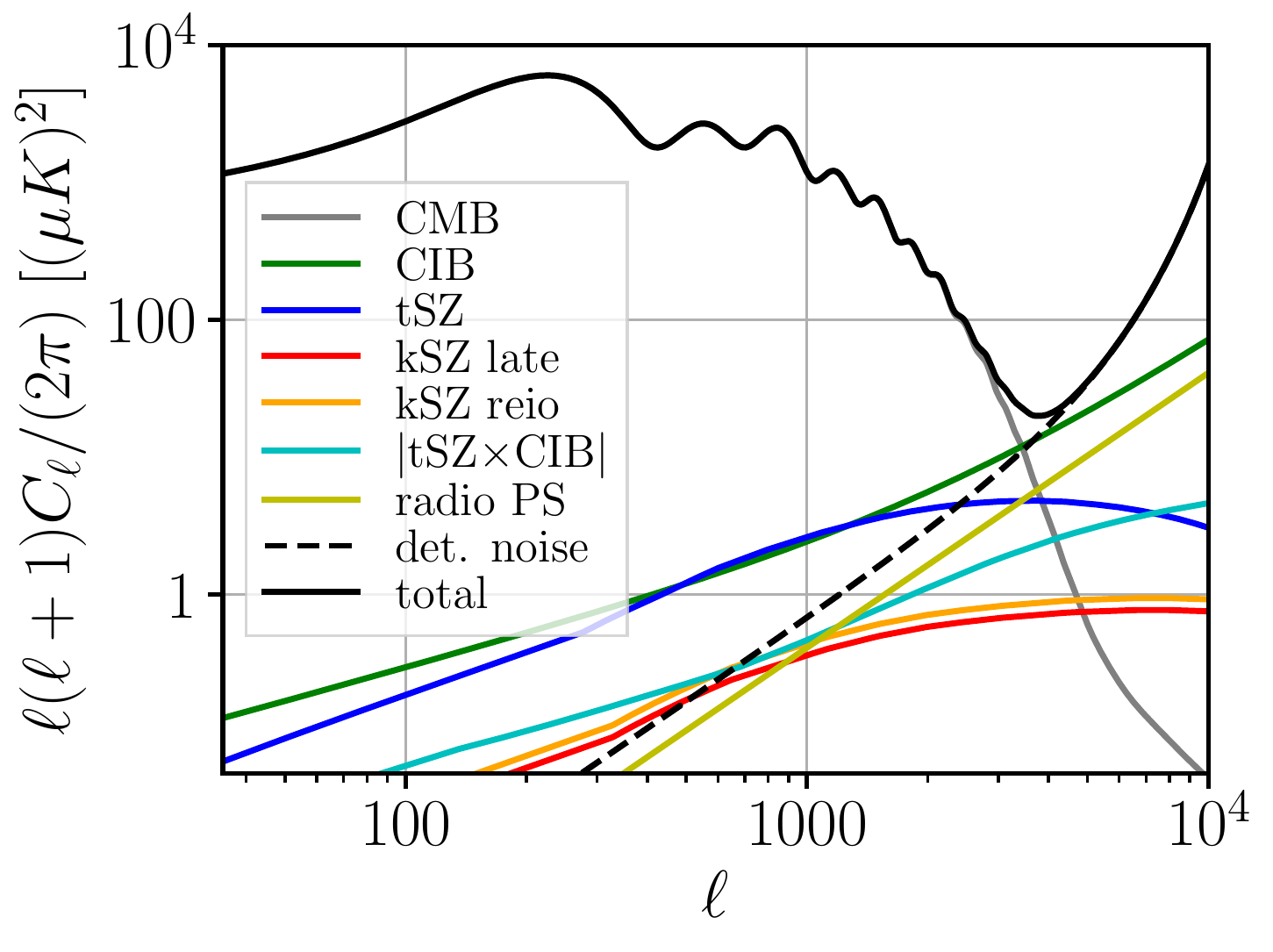}
\caption{
Extragalactic foreground power spectra at 143GHz from \cite{2013JCAP...07..025D}, compared to the lensed CMB and the detector noise ($7 \mu$K' white noise, $1.4'$ beam FWHM at 143 GHz).
A point source cut of $15$ mJy is assumed for radio PS, CIB and tSZ sources.
In the absence of foreground cleaning, the foregrounds are larger than the detector noise on scales $\ell\leq 3500$, and comparable to the lensed CMB at $\ell=3500$, where most of the lensing information comes from.
}
\label{fig:cl_foregrounds}
\end{figure}

Throughout this study, we use the model from \cite{2013JCAP...07..025D} for the power spectra of the extragalactic foregrounds at $143$ GHz, as shown in Fig.~\ref{fig:cl_foregrounds}.
This model includes a point source mask for all objects with flux larger than $15$ mJy.

In order to quantify the bias from lensed foregrounds to cross-correlations of CMB lensing with tracers, we also consider the LSST gold sample of galaxies, as described in the LSST Science Book \cite{2009arXiv0912.0201L}, chapter 3 and 13, with limiting magnitude in the i-band of $i_\text{lim} = 25.3$.
This galaxy sample contains $n_\text{gal} = 46 \times 10^{0.31*(i_\text{lim}-25)}$ galaxies per squared arcminute,
has bias $b(z) = 1 + 0.84 z$,
and redshift distribution
$dn/dz = n_\text{gal}  (z/z_0)^2  e^{-z/z_0} / (2 z_0)$,
where $z_0 \equiv 0.0417 i_\text{lim} - 0.744$.

\subsection{Effective redshift distribution of foreground sources}

Extragalactic foregrounds sources are associated with galaxies and clusters over a range of redshift.
Intuitively, the foreground redshift distribution relevant for lensing should be related to the galaxy or cluster redshift distributions.
Since the lensing estimators we consider are quadratic in the temperature map,
the relevant quantity should therefore be related to the redshift distribution of the (unlensed) foreground power spectrum $dC_\ell^f/dz$.
However, this quantity depends on the multipole $\ell$ considered:
generically, the large-scale power is mostly dominated by lower redshifts, because the same objects at lower redshift subtend a larger angle on the sky.
As a result, the relevant source redshift distribution for foreground lensing is scale-dependent, unlike in galaxy lensing where it is simply the galaxy $dn/dz$.
More precisely, as shown in App.~B of \cite{2018PhRvD..97l3539S}, the CMB lensing estimators at multipole $L$ respond to the following effective convergence:
\beq
\kappa^f_{\vL} =  \int d\chi \; W^{\kappa_f}(\chi, \vL) \; \delta_m(\vec{k} = \vL/\chi, \chi),
\eeq
where the effective lensing kernel $W^{\kappa_f}(\chi, \vL)$ depends on the multipole $L$ and is defined as:
\beq
W^{\kappa_f} (\chi, \vL) = \; \int d\chi_S
W^{f} (\chi_S, \vL) \; 
W^\kappa(\chi, \chi_S),
\label{eq:lensed_fg_kernel}
\eeq
where $W^\kappa(\chi, \chi_S)$ is the usual lensing kernel for a source at distance $\chi_S$, defined in the previous section, and the effective foreground redshift distribution $W^f (\chi_S, \vL)$ is
\beq
W^f (\chi_S, \vL)
=
\frac{dz_S}{d\chi_s} 
N_\vL
\int \frac{d^2\vl}{(2\pi)^2}\;
F_{\vl, \vL-\vl} f^{f,z}_{\vl, \vL-\vl},
\label{eq:foreground_dndz}
\eeq
with
\beq
f^{f,z}_{\vl_1, \vl_2}
\equiv
\left(
\frac{2 \vL}{L^2}
\right)
\cdot
\left[
\vl_1  \frac{d C^{f}_{\ell_1}}{dz_S}
+
\vl_2  \frac{dC^{f}_{\ell_2}}{dz_S}
\right].
\eeq
This is similar to galaxy lensing, where the lensing kernel is an integral over the source distribution. However, here, the source distribution is scale dependent.
As expected, Eq.~\eqref{eq:foreground_dndz} involves the redshift distribution $\frac{d C^{f}_{\ell}}{dz_S}$ of the unlensed foreground power spectrum. 
One important consequence is that the noise and resolution of the CMB experiment, which determine the lensing weights $F_{\vl, \vL-\vl}$, also determine how the $\frac{d C^{f}_{\ell}}{dz_S}$ terms are weighted to produce the effective foreground redshift distribution.
For instance, for a higher resolution CMB experiment, the CMB lensing reconstruction relies on smaller scales, thus upweighting $\frac{d C^{f}_{\ell}}{dz_S}$ at high $\ell$, which typically come from higher redshift sources.

\begin{figure}[h!!!!]
\includegraphics[width=0.45\textwidth]{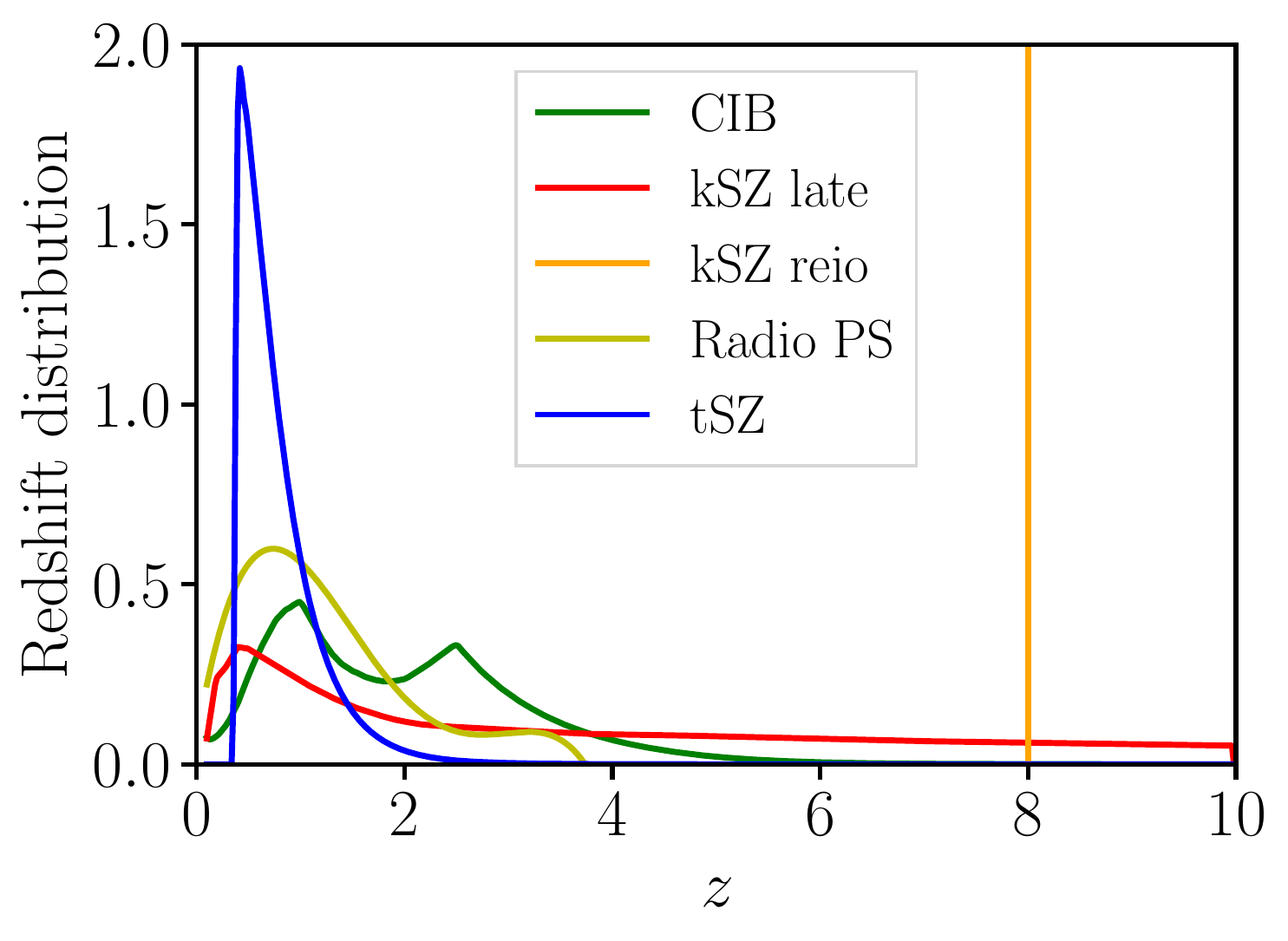}
\includegraphics[width=0.45\textwidth]{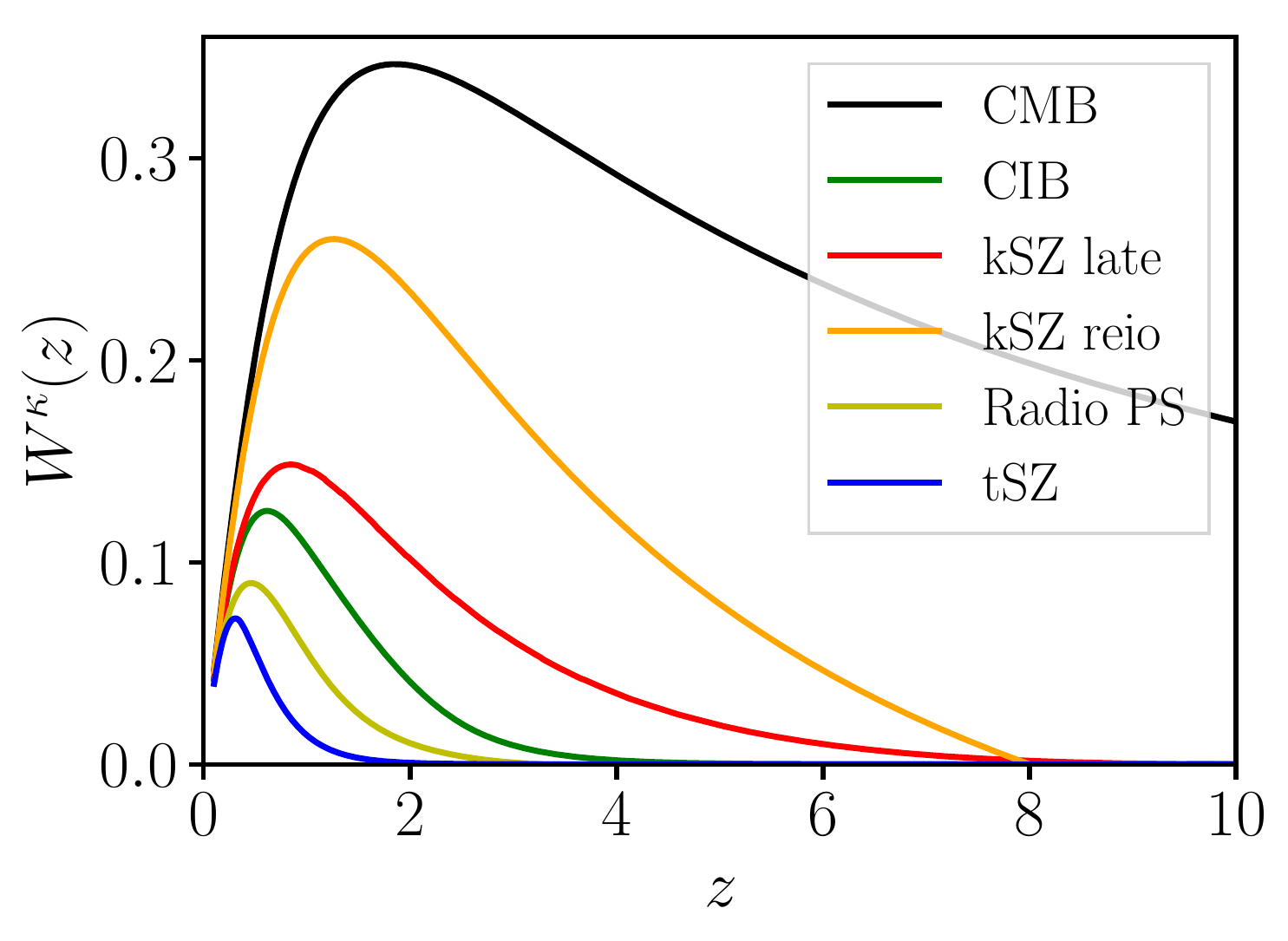}
\caption{
Extragalactic foregrounds are themselves lensed, because they are emitted at cosmological distances.
The exact foreground redshift distributions (Eq.~\eqref{eq:foreground_dndz}) depend on the experimental configuration (resolution, sensitivity, masking, etc.) and the lensing multipole $L$.
We approximate these redshift distributions as $W^f (\chi_S, \vL) \propto \frac{dC^f_{\ell=3000}}{d\chi_S}$ (left panel).
From these approximate redshift distributions, we infer the corresponding lensing kernels (right panel) from Eq.~\eqref{eq:lensed_fg_kernel}, showing which redshifts contribute to the foreground lensing convergence.
These lensing kernels determine the amplitude of the foreground lensing convergence and its correlation with the CMB lensing convergence.
Our approximations make these foreground lensing kernels independent of the experimental configuration and lensing multipole $L$.
}
\label{fig:redshift_distribution_foregrounds}
\end{figure}

For the CMB S3 experiment we consider, and assuming $l_\text{max T} = 3500$, most of the CMB lensing signal-to-noise comes from temperature multipoles $\ell \sim 3000$.
We therefore ignore the $L$-dependence of the foreground lensing kernels, and approximate them as
$W^f (\chi_S, \vL) \propto \frac{dC^f_{\ell=3000}}{d\chi_S}$.
The problem then simplifies to modeling 
$\frac{dC^f_{\ell=3000}}{d\chi_S}$
for each foreground of interest.
For the CIB, we use the value computed in \cite{2018PhRvD..97l3539S}, following the CIB halo model from \cite{2012A&A...537A.137P}, using the luminosity functions from \cite{2012ApJ...757L..23B, 2013A&A...557A..66B}.
For tSZ, we implement the halo model in \cite{2013PhRvD..88f3526H}.
For the late time kSZ, we use the redshift distribution from Fig.~6 in \cite{2012ApJ...756...15S}, corresponding to their L60CSFz2 model.
For the reionization kSZ, we assume a single source redshift at $z=8$, consistent with the redshift of a step-like reionization in (\cite{2018arXiv180706209P} table A1).
For the radio PS, we adopt the redshift distribution Eq.~26 in \cite{2010A&ARv..18....1D}, describing the source sample from \cite{2008MNRAS.385.1297B}, selected in NVSS at 1.4GHz. Because the frequency dependence of the synchrotron emission is very mild, sources at 1.4GHz and 143GHz mostly coincide, and correspond to active galactic nuclei.
These approximate effective redshift distributions are shown in Fig.~\ref{fig:redshift_distribution_foregrounds}, along with the corresponding foreground lensing kernels.
Our knowledge of the foreground redshift distributions is somewhat uncertain.
For this reason, the foreground lensing biases we obtain are also uncertain, and should be considered reasonable values rather than exact values.

\subsection{Foreground lensing power spectra}

\begin{figure}[h!!!!]
\includegraphics[width=0.45\textwidth]{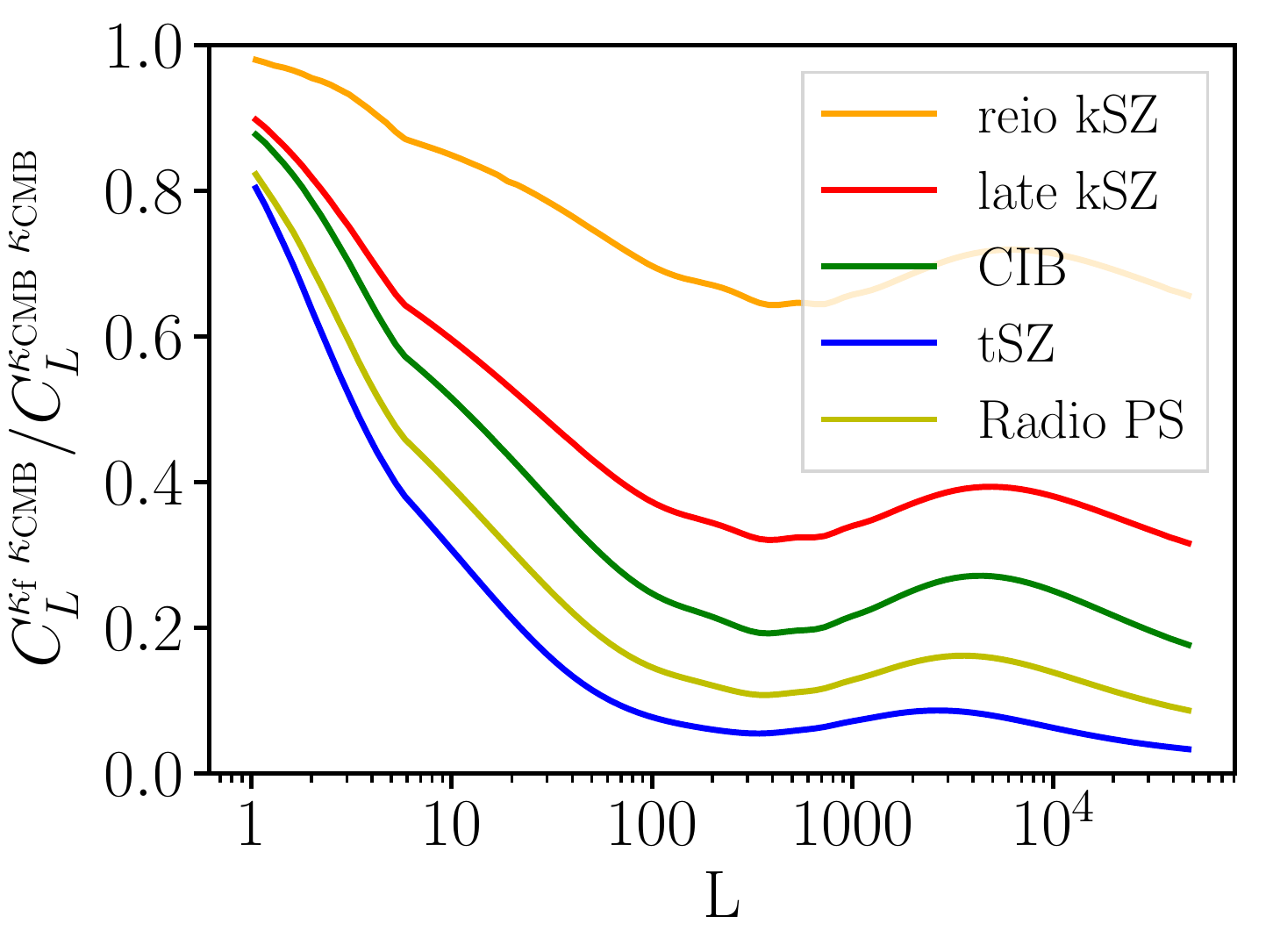}
\includegraphics[width=0.45\textwidth]{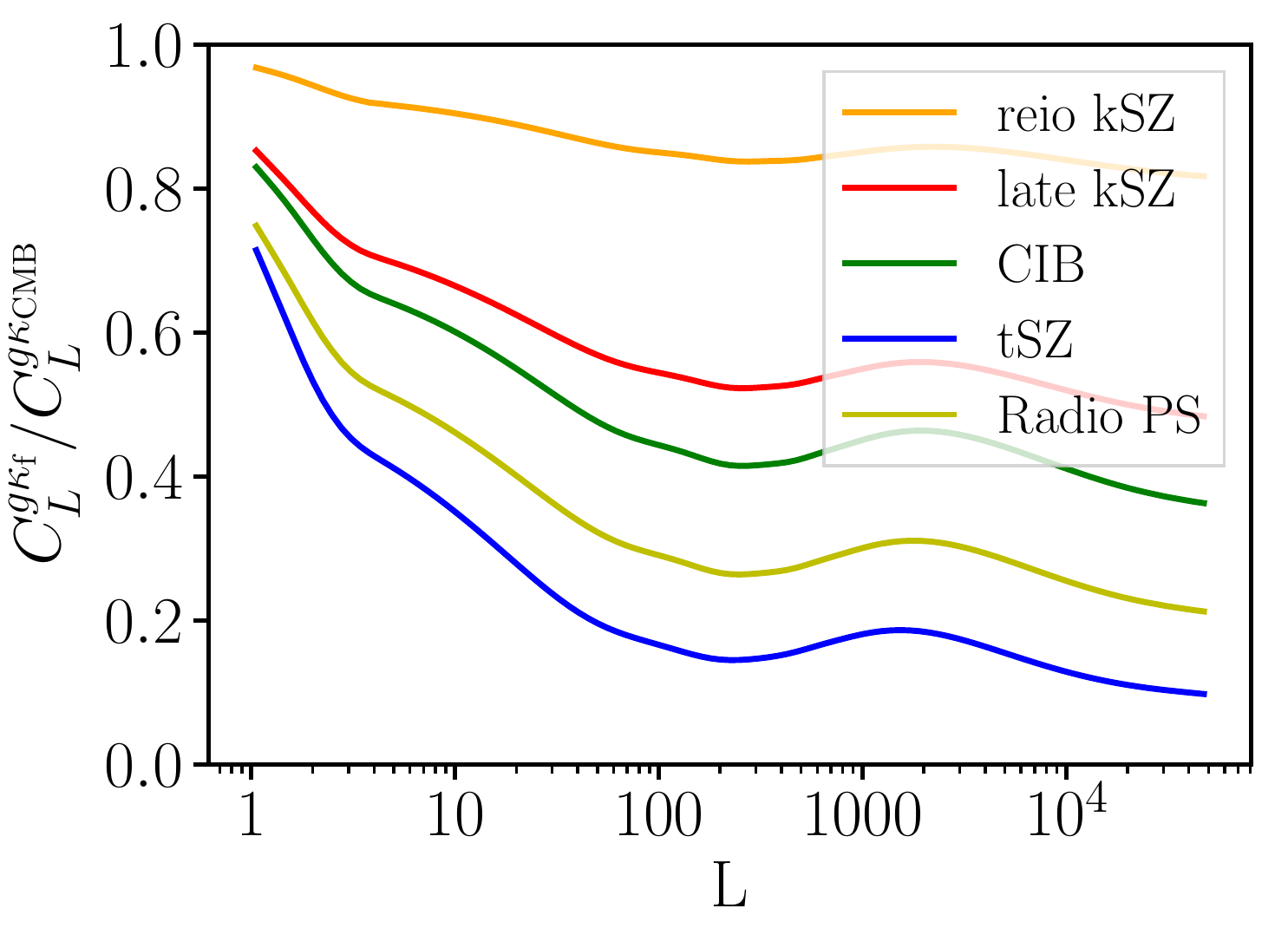}
\caption{
Since foreground sources lie at lower redshift than the CMB, the lensing they experience is less than that of the CMB, although of the same order of magnitude.
The relevant reduction factor is $C_L^{\kappa_f\kappa_\text{CMB}} / C_L^{\kappa_\text{CMB}\kappa_\text{CMB}}$ for the primary bias on the CMB lensing auto-spectrum (left panel, from Eq.~\eqref{eq:primary_bias}) and $C_L^{g\kappa_f} / C_L^{g\kappa_\text{CMB}}$ for the bias in cross-correlation with LSST galaxies (right panel, from Eq.~\eqref{eq:bias_cross}).
}
\label{fig:foreground_lensing_reduction}
\end{figure}

We use again the Limber and flat sky approximations, and obtain the auto and cross-spectra of the foreground lensing convergences:
\beq
C_L^{\kappa_1 \kappa_2}
=
\int \frac{d\chi}{\chi^2} \;
W^{\kappa_1}(\chi) W^{\kappa_2}(\chi)
P_m (k = \frac{L+1/2}{\chi}, z(\chi)).
\eeq
How small is foreground lensing compared to CMB lensing?
As we demonstrate in the next section, the relevant quantity to assess the bias in CMB lensing auto-spectrum is 
$C_L^{\kappa_f \kappa_\text{CMB}} / C_L^{\kappa_\text{CMB} \kappa_\text{CMB}}$, 
shown in the left panel of Fig.~\ref{fig:foreground_lensing_reduction}.
To assess the bias in cross-correlation with a tracer $g$, the relevant quantity is $C_L^{\kappa_f g} / C_L^{\kappa_\text{CMB} g}$,
shown in the right panel of Fig.~\ref{fig:foreground_lensing_reduction} for the LSST gold sample.
These plots show that the amplitude of foreground lensing for $L=100-1000$ ranges between $5\%$ (for tSZ) and $85\%$ (for the reionization kSZ) of that of CMB lensing, depending on the foreground considered.

\section{Bias to CMB lensing from lensed foregrounds}

In this section, we use the foreground power spectra and redshift distributions from the previous section to predict the bias to CMB lensing from lensed foregrounds.
Throughout, we make the simplifying approximation that each foreground map $f$ is lensed by one lensing convergence field $\kappa_f$, computed from the redshift distribution of the foreground.
In reality, each redshift slice of the foreground is lensed separately by the convergence field corresponding to that source redshift.
We believe that this approximation is sufficient, given the uncertainty in the foreground redshift distributions, and leave its exploration to future work.

Furthermore, since the lensing convergence and the foregrounds are both produced by the matter distribution in the universe, they are correlated.
The correlation between foregrounds and the CMB lensing convergence is the origin of the usual foreground biases to CMB lensing \cite{2014JCAP...03..024O, 2014ApJ...786...13V, 2018PhRvD..97b3512F, 2019PhRvL.122r1301S}, and we shall therefore not discuss it further.
The correlation between foregrounds and the foreground lensing convergence contributes additional bias in principle. However, this will be smaller than the previous one, and we shall only mention it in this section.

We compute all the bias terms described in this section in two independent ways, as described in App.~\ref{sec:ana_num_methods}, and find a good agreement everywhere.

\subsection{Response of CMB lensing estimators to lensed contaminants}

Following Sec.~VIII in \cite{2018PhRvD..97l3539S}, any quadratic combination of lensed foreground maps effectively constitutes a (potentially biased and suboptimal) foreground lensing estimator.
In particular, the CMB lensing estimators of the form Eq.~\eqref{eq:quadratic_estimator} are such quadratic combinations.
As a result, they generically have non-zero response to the foreground convergence $\kappa_f$.
Indeed, when applied to a lensed foreground map $f$, the quadratic estimators above have the following response to foreground lensing:
\beq
\langle \mathcal{Q}_\vL \left[ f, f \right] \rangle
=
\mathcal{R}^f_\vL \kappa^f_\vL,
\text{ with }
\mathcal{R}^f_\vL
=
\frac{\int \frac{d^2 \vl}{(2\pi)^2} F_{\vl, \vL-\vl} \; f^f_{\vl, \vL-\vl}}
{\int \frac{d^2 \vl}{(2\pi)^2} F_{\vl, \vL-\vl} \; f^\text{CMB}_{\vl, \vL-\vl}},
\label{eq:lensing_response}
\eeq
where the expectation value is at fixed convergence field,  varying the realization of the unlensed foreground map, like in the derivation of the standard quadratic estimator \cite{2002ApJ...574..566H}.

\begin{figure}[h!!!!]
\includegraphics[width=0.95\textwidth]{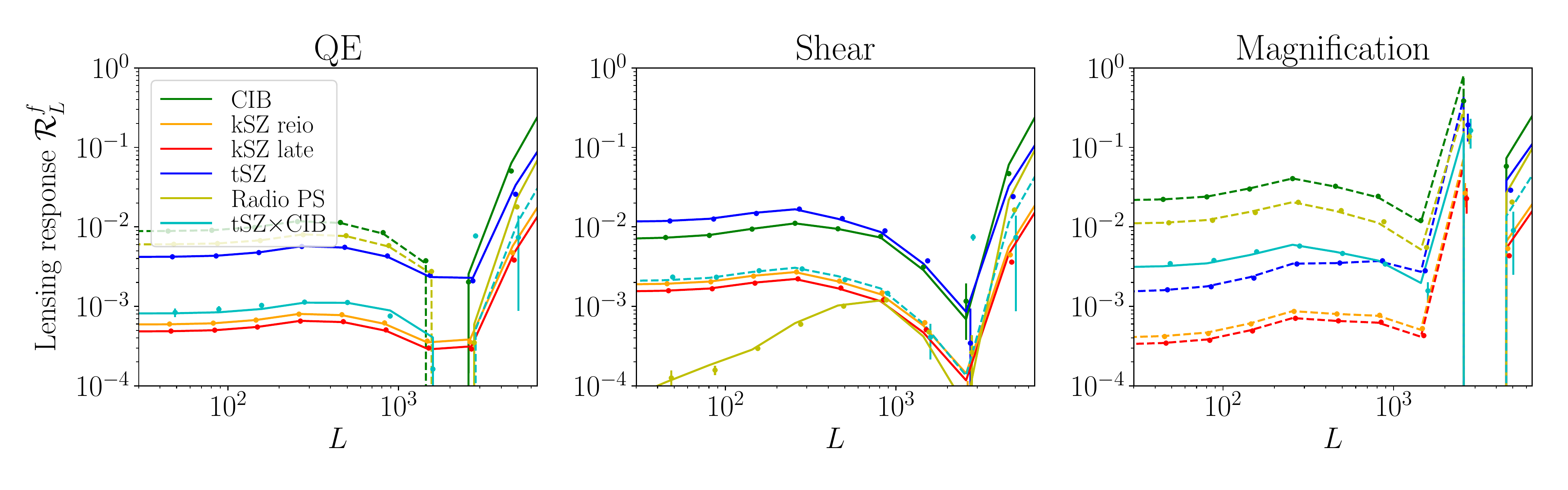}
\caption{
The lensing responses $\mathcal{R}_\vL^f$ from Eq.~\eqref{eq:lensing_response} for the QE (left), shear (center), and magnification (right) estimators
show what fraction of the foreground lensing convergence leaks into the CMB lensing estimator.
A response of unity would mean that the CMB lensing estimator has a bias equal to one times the foreground lensing convergence.
In practice, the responses for the various foregrounds and estimators are found to be of order percent for $L\lesssim 1000$.
The sign varies depending on the multipole, the foreground and the estimator (dashed lines represent negative responses).
The points with error bars are the simulation results, and the lines are the analytical calculations, binned like the simulations. The points and curves for each foreground are slightly shifted horizontally to improve the visibility of error bars.
}
\label{fig:lensing_responses}
\end{figure}

If the unlensed power spectra of the foreground and CMB were identical, the response $\mathcal{R}^f_\vL$ would be unity on all scales, and the CMB lensing estimator would also be an unbiased foreground lensing estimator.
If the foreground power spectrum $C^{f}$ is reduced by a factor $\alpha$, e.g., from multi-frequency component separation, then the response $\mathcal{R}_\vL^f$ is also reduced by the same factor $\alpha$.
It is thus clear that the value of the response $\mathcal{R}^f_\vL$ depends both on the amplitude and shape of the foreground power spectrum $C^{f}$, and can in principle take any value and sign.
In practice, Fig.~\ref{fig:lensing_responses} shows that $\mathcal{R}_\vL^f$ is of order one percent for $L\lesssim 1000$, with different signs depending on the foreground and choice of estimator (QE, shear, magnification).
The shear estimator does not reduce the response to foreground lensing compared to the QE. 
This is not surprising since foreground lensing is a true lensing effect, which the shear is built to include.
Furthermore, the lensing responses for shear and magnification have opposite signs. As a result, comparing shear and magnification provides a useful null test to detect the presence of lensed foreground bias.

As a result, if a lensed foreground is present in the temperature map, the quadratic estimator will inevitably partially reconstruct the lensing of the foreground:
\beq
\langle \mathcal{Q}_\vL \left[ T+f, T+f \right] \rangle
=
\kappa^\text{CMB}_\vL
+
\mathcal{R}_\vL^f \kappa^f_\vL
+\mathcal{O}\left( \kappa_\text{CMB}^2, \kappa_f^2, \kappa_\text{CMB} \kappa_f \right)
,
\eeq
where again the expectation value is at fixed lensing fields ($\kappa^\text{CMB}, \kappa^f$) but marginalizing over the unlensed fields ($T^0, f^0$).
This term produces biases in CMB lensing auto and cross-correlation, as we explain below.

Furthermore, two distinct unlensed foreground components $f$ and $f'$ may have a significant correlation.
This is the case for example for CIB and tSZ.
In this case, additional biases to CMB lensing occur, coming from terms of the form
\beq
\mathcal{Q}_\vL \left[ f, f' \right]
+
\mathcal{Q}_\vL \left[ f', f \right]
=
\mathcal{R}_\vl^{ff'} \left[ \kappa_\vL^{f}
+ \kappa_\vL^{f'} \right],
\eeq
where the additional response $\mathcal{R}_\vl^{ff'}$ is computed as in Eq.~\eqref{eq:lensing_response}, except that $f^f$ is replaced by $f^{ff'}$.
This response is typically smaller, reduced by a factor of order the correlation coefficient between $f$ and $f'$.
We quantify these extra terms below, in the case of CIB and tSZ.

\subsection{Bias in cross-correlation}

The response of CMB lensing quadratic estimators to the foreground convergence naturally leads to a bias in cross-correlation with any tracer $g$ (e.g., galaxy or cluster number density, or galaxy shear):
\beq
\frac{\delta C_L^{\kappa_\text{CMB} g}}{C_L^{\kappa_\text{CMB} g}}
=
\mathcal{R}_\vL^f \;\;
\frac{C_L^{\kappa_f g}}{C_L^{\kappa_\text{CMB} g}}.
\label{eq:bias_cross}
\eeq

\begin{figure}[h!!!!]
{\centering
\includegraphics[width=0.5\textwidth]{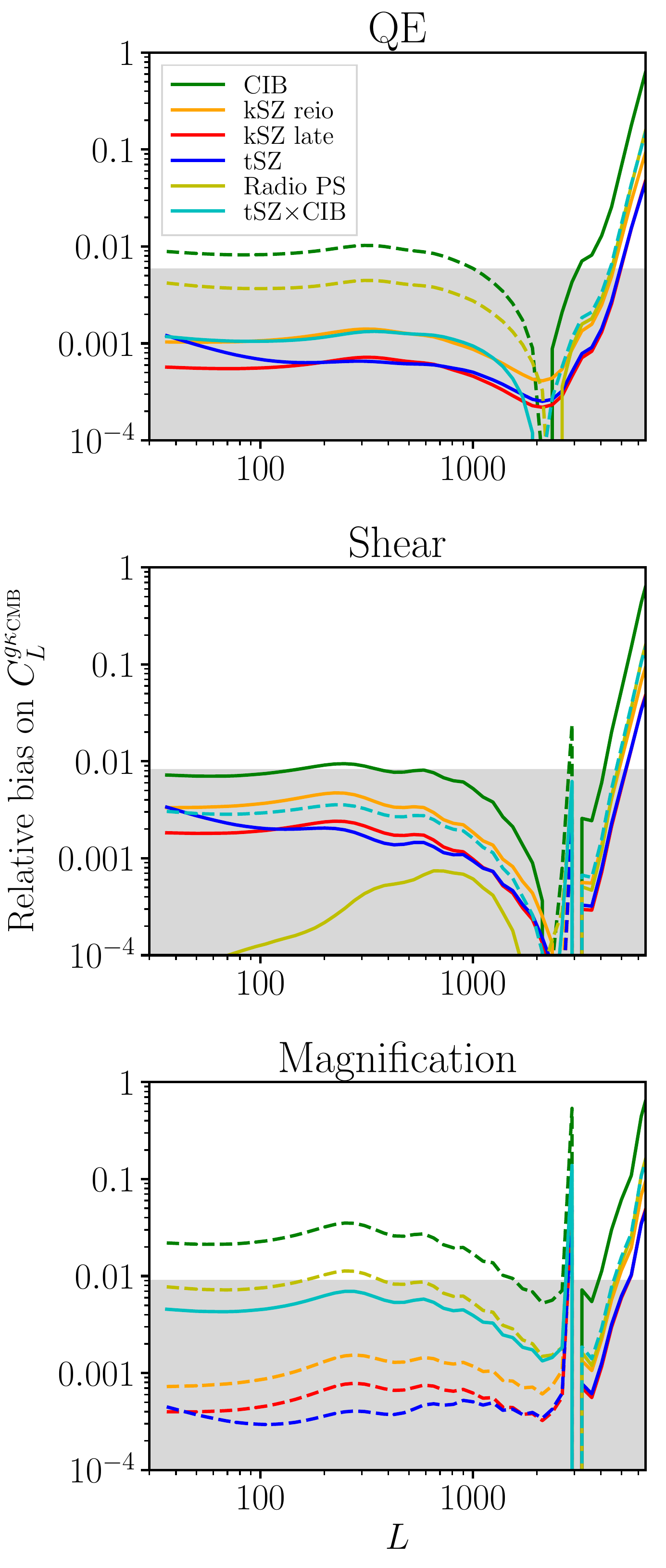}
}
\caption{
The relative bias from lensed foregrounds on the cross correlation of CMB lensing from Simons Observatory and the LSST gold galaxies is one percent or larger, 
depending on the foreground, when using the QE (top), shear (center) and magnification (bottom) estimators. 
This percent-level systematic bias is larger than the statistical uncertainty (grey shaded areas) of $0.6\%, 0.8\%$ and $0.9\%$ on the amplitude of the cross-correlation for the QE, shear and magnification respectively (assuming $f_\text{sky}=40\%$, $L_{\text{max, }\kappa}=1000$ and including cosmic variance).
Dashed lines represent negative biases.
The spikes at $L\simeq 3000$ for the shear and magnification estimators are due to spikes in their noise power spectra at this multipole, as shown in Fig.~2 of \cite{2019PhRvL.122r1301S}.
}
\label{fig:bias_cross}
\end{figure}

When two correlated foregrounds are present, such as CIB and tSZ, an additional bias from the cross term arises:
\beq
\frac{\delta C_L^{\kappa_\text{CMB} g}}{C_L^{\kappa_\text{CMB} g}}
=
\mathcal{R}_\vL^{ff'} \;\;
\frac{\left[C_L^{\kappa_f g} + C_L^{\kappa_{f'} g}\right]}{C_L^{\kappa_\text{CMB} g}}.
\eeq
Fig.~\ref{fig:bias_cross} shows that these biases are typically of order one percent of the cross-power spectrum $C_L^{\kappa_\text{CMB} g}$,
i.e. significantly larger than the statistical uncertainty (lensing reconstruction noise plus cosmic variance) for Simons Observatory lensing and the LSST gold galaxy sample.

\subsection{Bias in auto-correlation}

In auto-correlation, several bias terms arise.
These are derived by expanding 
$\langle \mathcal{Q} \left[ T+f, T+f \right] \mathcal{Q} \left[ T+f, T+f \right] \rangle$,
using the bilinearity of the quadratic estimators.

\begin{figure}[h!!!!]
\includegraphics[width=0.95\textwidth]{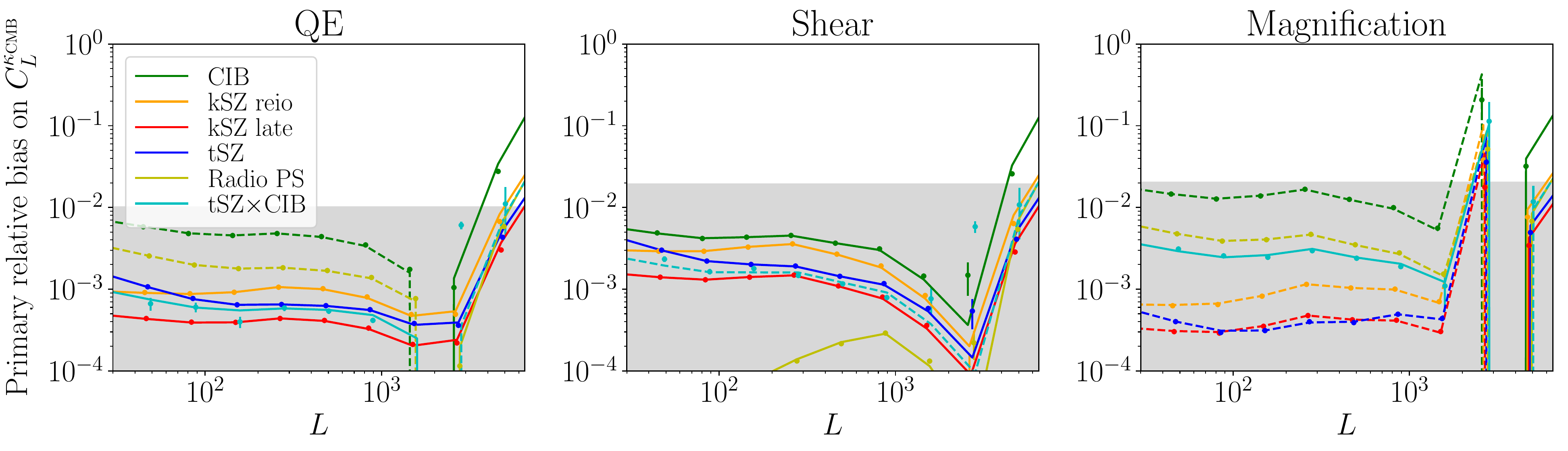}
\caption{
The primary relative bias from lensed foregrounds on the power spectrum of CMB lensing from Simons Observatory is of order one percent or less, when using the QE (left), shear (center) and magnification (right) estimators.
This percent-level systematic bias can be as large as the statistical uncertainty (grey shaded areas) of $1\%, 2\%$, and $2\%$ on the amplitude of the CMB lensing power spectrum for the QE, shear and magnification (assuming $f_\text{sky}=40\%$ and including cosmic variance).
Dashed lines represent negative biases.
The points with error bars are the simulation results, and the lines are the analytical calculations, binned like the simulations. The points and curves for each foreground are slightly shifted horizontally to improve the visibility of error bars.
}
\label{fig:primary_bias}
\end{figure}

We call ``primary'' bias the terms where one quadratic estimator is applied to two foreground maps and the other is applied to two CMB maps:
$\langle \mathcal{Q} \left[ f, f \right] \mathcal{Q} \left[ T, T \right] \rangle + f\leftrightarrow T$.
These terms are analogous to the cross-correlation case, with an additional combinatorial factor $2$:
\beq
\text{\textbf{Primary bias: }}
\frac{\delta C_L^{\kappa_\text{CMB} \kappa_\text{CMB}}}{C_L^{\kappa_\text{CMB} \kappa_\text{CMB}}}
=
2\; \mathcal{R}_\vL^f \;\;
\frac{C_L^{\kappa_f \kappa_\text{CMB}}}{C_L^{\kappa_\text{CMB} \kappa_\text{CMB}}}.
\label{eq:primary_bias}
\eeq
Again, when two correlated foregrounds like CIB and tSZ are present, an additional primary bias arises from the cross term:
\beq
\frac{\delta C_L^{\kappa_\text{CMB} \kappa_\text{CMB}}}{C_L^{\kappa_\text{CMB} \kappa_\text{CMB}}}
=
2\; \mathcal{R}_\vL^{ff'} \;\;
\frac{\left[C_L^{\kappa_f \kappa_\text{CMB}} + C_L^{\kappa_{f'} \kappa_\text{CMB}} \right]}{C_L^{\kappa_\text{CMB} \kappa_\text{CMB}}}.
\eeq
These terms are shown to be percent level biases in the CMB lensing auto-spectrum in Fig.~\ref{fig:primary_bias}.

\begin{figure}[h!!!!]
\includegraphics[width=0.97\textwidth]{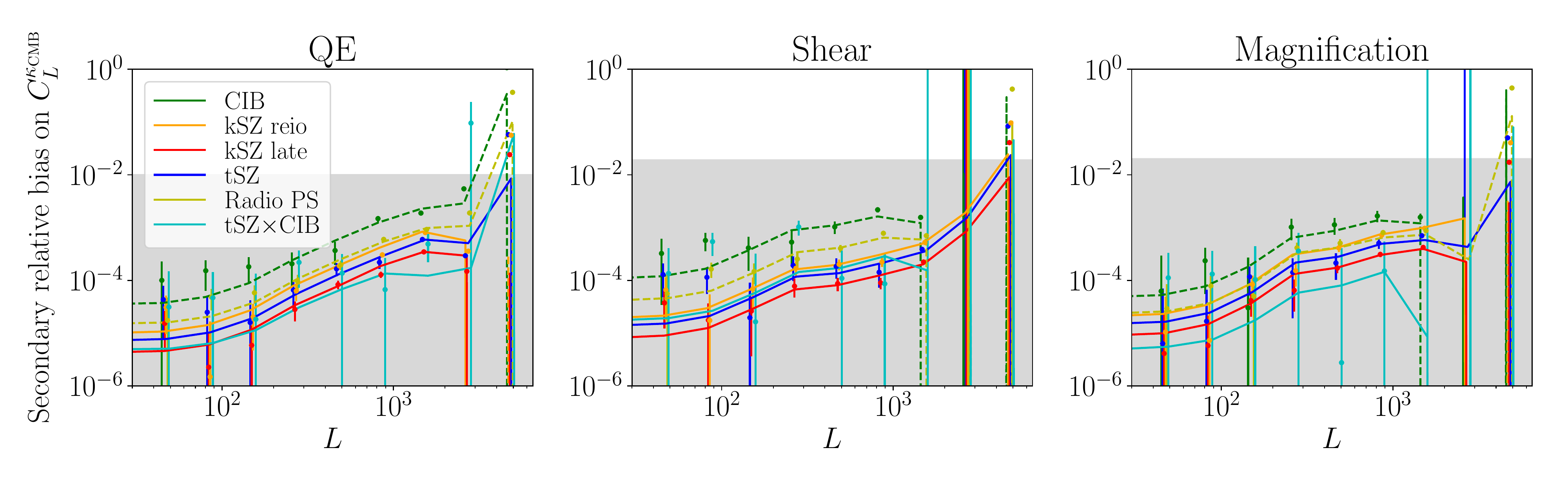}
\caption{
The secondary relative bias from lensed foregrounds on the power spectrum of CMB lensing is suppressed at low lensing multipoles due to the cancellation in Eq.~\eqref{eq:secondary_bias}.
However, for lensing multipoles of a few thousand, the secondary bias dominates over the primary bias.
The various panels show this for the QE (left), shear (center) and magnification (right) estimators.
The grey shaded areas represent the statistical uncertainty on the amplitude of the CMB lensing power spectrum ($1\%, 2\%$ and $2\%$ for the QE, shear and magnification, assuming $f_\text{sky}=40\%$ and including cosmic variance), and dashed lines represent negative terms.
The points with error bars are the simulation results, and the lines for QE are the analytical calculations, evaluated (not binned) at the simulation points. The lines for shear and magnification are simply connecting the points, not analytical calculations. The points and curves for each foreground are slightly shifted horizontally to improve the visibility of error bars.
}
\label{fig:sec_bias}
\end{figure}

In auto-correlation, an additional ``secondary'' bias is present, due to the terms where each quadratic estimator is applied to one CMB map and one foreground map:
$\langle \mathcal{Q} \left[ T, f \right] \mathcal{Q} \left[ T, f \right] \rangle + \text{perm.}$
.
The corresponding secondary bias is derived in App.~\ref{sec:secondary_bias}:
\beq
\bal
\text{\textbf{Secondary bias: }}
\delta C_{L_0}^{\kappa_\text{CMB} \kappa_\text{CMB}}
&=
8 N_\vL^2
\int\frac{d^2 \vL}{(2\pi)^2}
\int\frac{d^2 \vl}{(2\pi)^2}
F_{\vl, \vL_0-\vl}
F_{\vl-\vL-\vL_0, \vL-\vl}
\alpha_{\vL, \vl-\vL}
C^{\kappa_\text{CMB} \kappa_f}_{\vL}\\
&\times\left[ 
\alpha_{-\vL, \vl-\vL_0}
C^{\text{CMB}}_{\vL_0-\vl}
C^{f}_{\vl-\vL}
+
\alpha_{-\vL, \vL_0+\vL-\vl}
C^{\text{CMB}}_{\vl-\vL}
C^{f}_{\vl-\vL-\vL0}
\right]\\
\eal
\label{eq:secondary_bias}
\eeq
where we used the notation $\alpha_{\vL, \vl-\vL} = -2 \frac{\vL}{L^2} \cdot \left( \vl-\vL \right)$, and $C^\text{CMB}$ and $C^f$ represent the unlensed CMB and foreground power spectra.
Again, when correlated foregrounds such as CIB and tSZ are present, an additional secondary bias appears, obtained by  substituting $C^{ff'}$ to $C^f$ and $\left[ C^{\kappa_\text{CMB} \kappa_f} + C^{\kappa_\text{CMB} \kappa_{f'}} \right]$
to $C^{\kappa_\text{CMB} \kappa_f}$ in Eq.~\eqref{eq:secondary_bias}.
Na\"ively, this secondary bias should be of the same order of magnitude as the primary bias.
However, in the limit $L, L_0 \ll \ell$, i.e. when reconstructing large-scale lensing modes from small-scale temperature modes, the terms in the square bracket cancel exactly.
As a result,  the secondary bias is negligible compared to the primary bias at low lensing multipoles.
However, the secondary bias dominates over the primary bias at lensing multipoles of a few thousand.
This is shown in Fig.~\ref{fig:sec_bias}.
\begin{figure}[h!!!!]
\includegraphics[width=0.5\textwidth]{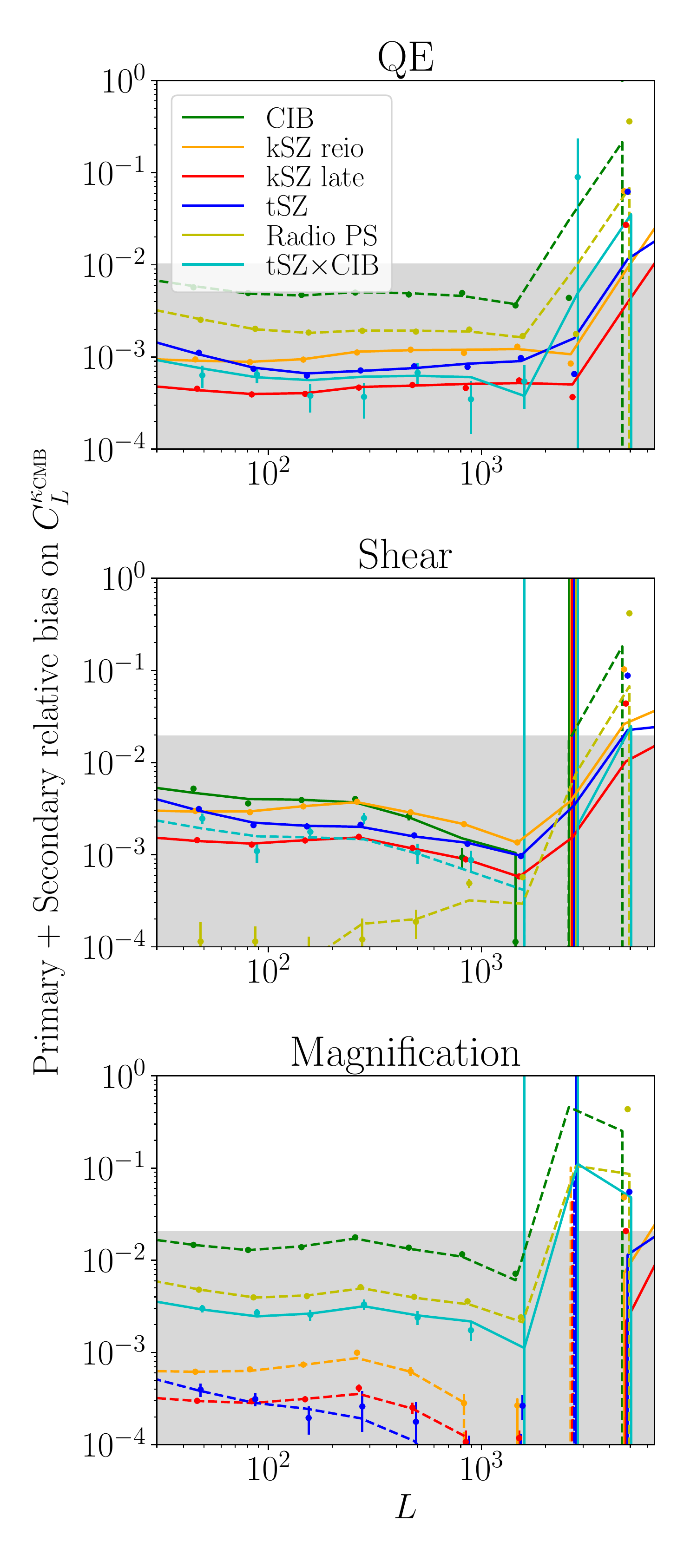}
\caption{
The total (primary + secondary) relative bias from lensed foregrounds on the power spectrum of CMB lensing from Simons Observatory is of order one percent, when using the QE (top), shear (center) and magnification (bottom) estimators.
It is dominated by the primary bias for $L\lesssim 1000$, then by the secondary bias on smaller scales.
The grey shaded areas represent the statistical uncertainty on the amplitude of the CMB lensing power spectrum ($1\%, 2\%$ and $2\%$ for the QE, shear and magnification, assuming $f_\text{sky}=40\%$ and including cosmic variance), and dashed lines represent negative terms.
The points with error bars are the simulation results, and the lines for QE are the analytical calculations, evaluated (not binned) at the simulation points. The lines for shear and magnification are simply connecting the points, not analytical calculations. The points and curves for each foreground are slightly shifted horizontally to improve the visibility of error bars.
The spikes at $L\simeq 3000$ for the shear and magnification estimators are due to spikes in their noise power spectra at this multipole, as shown in Fig.~2 of \cite{2019PhRvL.122r1301S}.
}
\label{fig:primsec_bias}
\end{figure}
Overall, the sum of primary and secondary lensed foreground biases constitutes a percent-level bias in CMB lensing auto-spectrum, comparable with the statistical uncertainty (lensing reconstruction noise plus cosmic variance) for a stage III CMB experiment.

Finally, a ``4-point'' bias is also present, where both quadratic estimators are evaluated on two foreground maps:
\beq
\text{\textbf{4-point bias: }}
\delta C_{L}^{\kappa_\text{CMB} \kappa_\text{CMB}}
=
\left(\mathcal{R}_L^{f} \right)^2 C_L^{\kappa_f \kappa_f} + N^{(1)}\text{-like term}.
\eeq
Since the response to foreground lensing is small for $L \lesssim 1000$, i.e. $\mathcal{R}_L^f \ll 1$, this 4-point bias is expected to be much smaller than the primary and secondary biases on these scales, and we will not discuss it further in this paper.
However, it may be large on smaller scales $L \gtrsim 1000$.

All the biases due to lensed foregrounds discussed so far are present whether or not the foreground component is a Gaussian random field.
This is in contrast with the non-Gaussian foreground biases usually discussed in the literature, which are caused by the non-Gaussianity of the foregrounds, and not the fact that they are lensed.
There exist additional biases due to the foregrounds being both non-Gaussian and lensed. To lowest order in $\kappa_f$, these terms are of the form:
\beq
\bal
\langle \mathcal{Q}[f_1, f_2] \mathcal{Q}[f_3, f_4] \rangle
\sim
\langle \kappa_{f_i} f_1 f_2 f_3 f_4 \rangle
&\sim \langle \kappa_{f_i} f_1 \rangle \langle f_2 f_3 f_4 \rangle 
\leftarrow \text{foreground bispectrum}\\
&+ \langle \kappa_{f_i} f_1 f_2 \rangle \langle f_3 f_4 \rangle
\leftarrow \text{lensing-foreground-foreground bispectrum}\\
&+ \langle \kappa_{f_i} f_1 f_2 f_3 f_4 \rangle_c
\leftarrow \text{connected 5-point function}
\eal
\eeq
We expect these terms to be smaller than the usual non-Gaussian foreground biases \cite{2013MNRAS.431..609N, 2014MNRAS.438.1507N, 2014ApJ...786...13V, 2014JCAP...03..024O, 2018PhRvD..97b3512F} by the foreground lensing response $\mathcal{R}_\vL^f$,
but we leave the evaluation of these terms to future work.

\section{Conclusion}

In this paper, we quantified for the first time the bias to CMB lensing auto and cross-correlations due to the presence of lensed foregrounds in the observed temperature map.
For an experiment similar to Simons Observatory,
and in the absence of multi-frequency foreground cleaning, this lensed foreground bias is a percent-level effect for both the CMB lensing power spectrum and for the cross-correlation of CMB lensing with LSST galaxies.
This bias is thus marginally significant in auto-correlation, and highly significant in cross-correlation.
For future polarization-dominated CMB lensing experiments like CMB S4, we expect the lensed foreground bias to be a lesser problem, since extragalactic foregrounds are expected to be smaller in polarization than in temperature.

For the standard quadratic estimator and the magnification estimators, the standard non-Gaussian foreground biases are typically more important than the lensed foreground bias.
Any method that successfully controls the former will thus automatically control the latter.
On the other hand, the shear estimator is mostly insensitive to the standard foreground bias. As a result, the lensed foreground bias is dominant. 
It needs to be reduced in order to provide an unbiased lensing measurement.

Any mitigation method that reduces the level of foregrounds in CMB temperature maps will also reduce the lensed foreground bias.
This is the case of multi-frequency component separation, scale cuts in the temperature map, masking or inpainting point sources (below the $15$ mJy assumed used in this paper), and the cleaned-gradient estimators.
This suggests that a combination of such methods may be the best approach.

On the other hand, mitigation methods that rely on the non-Gaussian structure of foregrounds will not in general reduce the lensed foreground bias, since the lensed foreground bias is present even if the foreground of interest is perfectly Gaussian.
Examples are the standard foreground bias hardening and the shear estimator.
Indeed, we have shown that the shear estimator, designed to distinguish the spatial symmetry of the lensing shear (quadrupole) from those of the non-Gaussian foregrounds (monopole), is sensitive to the lensed foreground bias.
However, the lensed foreground biases in shear and magnification have opposite signs. Comparing shear and magnification estimators therefore still provides a useful null test.

In App.~\ref{sec:bias_hardening}, we suggest a ``lensed foreground bias-hardening'' to reduce the lensed foreground bias.
Contrary to the usual bias hardening, this lensed foreground bias hardening does not assume any knowledge of the often uncertain non-Gaussianity of the foreground.
However, it relies on a knowledge of the power spectrum and source redshift distribution of the foreground.
The first estimator we derive has zero response to the lensing of a given foreground component, at the cost of an increased noise.
To avoid a potentially large noise cost,
we derive a second estimator, which instead minimizes the total variance from lensing noise plus residual lensed foreground bias.
We leave the exploration of these forms of bias hardening to future work.

Finally, one may remove the lensed foreground bias by subtracting their theory predictions, as computed in this paper.
An uncertainty on the theory prediction of the lensed foreground bias of order ten percent would be acceptable.
Quantifying the uncertainty in the foreground source distributions assumed in this paper would be useful to assess whether such accuracy is reachable.
Furthermore, our study makes an important simplifying approximation, by assuming that each foreground map $f$ is lensed as a whole by a single convergence field $\kappa_f$, determined by the foreground redshift distribution.
In reality, each redshift slice of the foreground emission is lensed separately by a slightly different convergence field, determined by the redshift of that slice.
We believe that the error due to this approximation is comparable to the uncertainty in the foreground redshift distributions.
A complete analysis of foreground biases to CMB lensing including both the effect of foreground non-Gaussianity and of realistic foreground lensing, would be a worthwhile endeavor. This could be achieved by lensing realistic non-Gaussian simulations such as the \texttt{Websky}\footnote{\url{https://mocks.cita.utoronto.ca/index.php/WebSky_Extragalactic_CMB_Mocks}} mocks \cite{2019MNRAS.483.2236S}.

Finally, we suggest that foreground lensing may be considered a signal rather than a bias \cite{2004NewA....9..173C, 2018PhRvD..97l3539S, 2019arXiv190502084F}.

\section{Acknowledgments}

We are indebted to Marcelo Alvarez for 
making this collaboration possible,
and to Sukhdeep Singh and Yu Feng for their help.
We warmly thank Simone Ferraro, Alex van Engelen and Uros Seljak for many insightful discussions which inspired this project.
We thank Rahul Datta for his insight on the nature and redshift distribution of radio point sources.
We thank Vanessa Boehm, Simone Ferraro, Colin Hill, Ben Horowitz, Srinivasan Raghunathan and Martin White for their comments on an earlier version of this paper.

ES is supported by the Chamberlain fellowship at Lawrence Berkeley National Laboratory.
 
This work used resources of the National Energy Research Scientific Computing Center, a DOE Office of Science User Facility supported by the Office of Science of the U.S. Department of Energy under Contract No. DE-AC02-05CH11231.


\bibliography{refs}

\appendix

\section{Derivation of the secondary bias to CMB lensing from lensed foregrounds}
\label{sec:secondary_bias}

We consider the auto-spectrum of $\mathcal{Q}_\vL\left[ T+f, T+f \right]$,
where $T$ now represents the lensed CMB and $f$ the lensed foreground.
The quadratic estimators are by definition bilinear in their arguments, and can thus be expanded.
We further Taylor expand $T=T^0+T^1$ and $f=f^0+f^1$ to first order in the lensing convergence as follows:
\beq
T_\vl
=
T^0_\vl
+
T^1_\vl
+\mathcal{O} \left( \kappa^2 \right),
\eeq
where
\beq
T^1_\vl = \int \frac{d^2 \vL}{(2\pi)^2}\;
\alpha_{\vL, \vl-\vL}\;
\kappa_\vL
T^0_{\vl-\vL}
\eeq
and 
$\alpha_{\vl_1, \vl_2} = -2 \frac{\vl_1}{\ell_1^2} \cdot \vl_2$.
These expressions are valid both for the lensed CMB and a lensed foreground.
Finally, we keep only the terms that are first order in $\kappa_f \times \kappa_\text{CMB}$.

We thus get:
\beq
\bal
\delta C_{L_0}^{\kappa_\text{CMB}}
&=
8\langle \mathcal{Q}_{\vL_0}\left[ f^0, f^1 \right] \, \mathcal{Q}_{\vL_0}\left[ T^0, T^1 \right]\rangle^\prime \\
&+ 
8\langle\mathcal{Q}_{\vL_0}\left[ T^0, f^0 \right] \, \mathcal{Q}_{-\vL_0}\left[ T^1, f^1 \right]\rangle^\prime \\
&+ 
8\langle\mathcal{Q}_{\vL_0}\left[ T^0, f^1 \right] \, \mathcal{Q}_{-\vL_0}\left[ T^1, f^0 \right]\rangle^\prime , \\
\eal
\eeq
where $\langle ... \rangle^\prime \equiv \langle ... \rangle / (2\pi)^2 \delta^D(\vec{0})$. 
The first line corresponds to the primary bias, 
and we refer to the additional terms on line 2 and 3 as secondary bias.
They can be expressed explicitly as:
\beq
\bal
8\langle\mathcal{Q}_{\vL_0}\left[ T^0, f^0 \right] \,
&
\mathcal{Q}_{-\vL_0}\left[ T^1, f^1 \right]\rangle
=\\
&
8 N_\vL^2
\int\frac{d^2 \vL}{(2\pi)^2}
\int\frac{d^2 \vl}{(2\pi)^2}
F_{\vl, \vL_0-\vl}
F_{\vl-\vL-\vL_0, \vL-\vl}
\alpha_{\vL, \vl-\vL}
\alpha_{-\vL, \vl-\vL_0}
C^{\text{CMB }0}_{\vL_0-\vl}
C^{f 0}_{\vl-\vL}
C^{\kappa^\text{CMB} \kappa^f}_{\vL}
\eal
\eeq
and
\beq
\bal
8\langle\mathcal{Q}_{\vL_0}\left[ T^0, f^1 \right] \, 
&
\mathcal{Q}_{-\vL_0}\left[ T^1, f^0 \right]\rangle
=\\
&
8 N_\vL^2
\int\frac{d^2 \vL}{(2\pi)^2}
\int\frac{d^2 \vl}{(2\pi)^2}
F_{\vl, \vL_0-\vl}
F_{\vL-\vl, \vl-\vL-\vL_0}
\alpha_{\vL, \vl-\vL}
\alpha_{-\vL, \vL_0+\vL-\vl}
C^{\text{CMB }0}_{\vl-\vL}
C^{f 0}_{\vl-\vL-\vL_0}
C^{\kappa^\text{CMB} \kappa^f}_{\vL}.
\eal
\eeq
Eq.~\eqref{eq:secondary_bias} is then obtained by adding these two terms together, and makes apparent their exact cancellation in the limit $L, L_0 \ll \ell$, where large-scale lensing modes are reconstructed from small-scale temperature modes.

\section{Comparison of analytical and numerical methods}
\label{sec:ana_num_methods}

\subsection{Analytical evaluation methods}

For the bias in CMB lensing cross-correlation, and for the primary and secondary biases to CMB lensing, we evaluate the response $\mathcal{R}_\vL^f$ from Eq.~\eqref{eq:lensing_response} by using the Fast Fourier Transform (FFT). Indeed, each integral is a sum of products and convolutions, which can be computed efficiently by successive products and FFT steps.
For the secondary bias, we use the python package \texttt{vegas}\footnote{\url{https://pypi.org/project/vegas/}} to compute the 4d integral with a Monte Carlo method.
All these calculations are performed in the publicly available repository \url{https://github.com/EmmanuelSchaan/LensedForegroundBias}, building upon the \texttt{ForQuE} module\footnote{\url{https://github.com/EmmanuelSchaan/ForQuE}}.

\subsection{Simulations}

In addition to the analytical calculations describe above, we also evaluate the foreground lensing responses, primary and secondary biases using simulated maps of the CMB, foregrounds and their lensing convergences. 
These calculations build upon the \texttt{LensQuEst} module\footnote{\url{https://github.com/EmmanuelSchaan/LensQuEst}}.
We generate Gaussian random fields (GRF) for the unlensed CMB $T^0$ and the unlensed foreground $f^0$.
We simulate the CMB lensing convergence $\kappa_\text{CMB}$ and foreground lensing convergence $\kappa_f$ as correlated GRF.
Using GRFs for these simulated maps is sufficient, since the lensed foreground biases we consider in this paper depend only on the foreground power spectra, and not on the detail of their non-Gaussian statistics.

We generate correlated GRFs as follows.
To generate GRF maps $m_1$ and $m_2$ with auto-spectra $C_L^{11}$ and $C_L^{22}$ and cross-spectrum $C_L^{12}$,
we first generate a GRF $m_1$ with power spectrum $C_L^{11}$, then define
$m_2 \equiv \alpha m_1 + m_\perp$,
where $\alpha_L = C_L^{12}/C_L^{11}$
and $m_\perp$ is a GRF with power spectrum 
$C_L^{22} - \left( C_L^{12} \right)^2 / C_L^{11}$.
This produces GRF maps $m_1$ and $m_2$ with the correct auto and cross-spectra.

At this point, we could simply lens the CMB and foreground maps with their respective convergences, add them, and apply the lensing estimators to the resulting map.
However, one would then need to perform $N^0$ subtraction, and the result would also be affected by a large noise from the lensing reconstruction.
Measuring the percent-level bias due to lensed foregrounds, with a precision of say a few percent, would then require a large amount of simulations.

Instead, we use the trick of lensing the CMB and foreground maps only to first order. In other words, we Taylor expand the lensed temperature map in powers of the convergence field: 
$T = T^0 + T^1 + \mathcal{O}\left( \kappa^2 \right)$,
where $T^0$ is the unlensed map, and $T^1\equiv \vec{\nabla}\phi\cdot \vec{\nabla}T^0$ is the first order lensing correction.
We perform the same operation for the foreground map $f$.

Applying the lensing estimators directly on the total map, CMB plus foreground, lensed to first order, would still require a $N^0$ subtraction, and would actually make the noise worse \cite{2016arXiv160303496A}.
Instead, we evaluate the lensing responses, primary and secondary biases  as follows:
\beq
\text{\textbf{Response: }}
\mathcal{R}_L^f
=
 \frac{\langle \left(\mathcal{Q}\left[ f_0, f_1 \right] + \mathcal{Q}\left[ f_1, f_0 \right]\right)  \,  \kappa_{f}\rangle}{\langle \kappa_{f} \, \kappa_{f}\rangle}
\label{eq:response_calc}
\eeq
\beq
\text{\textbf{Primary bias: }}
\frac{\delta C_L^{\kappa_\text{CMB} \kappa_\text{CMB}}}{C_L^{\kappa_\text{CMB} \kappa_\text{CMB}}}
=
2 \frac{\langle \left(\mathcal{Q}\left[ f_0, f_1 \right] + \mathcal{Q}\left[ f_1, f_0 \right]\right)  \,  \kappa_\text{CMB}\rangle}{\langle \kappa_\text{CMB} \, \kappa_\text{CMB}\rangle}
\label{eq:primary_calc}
\eeq
\beq
\bal
\text{\textbf{Secondary bias: }}
\frac{\delta C_L^{\kappa_\text{CMB} \kappa_\text{CMB}}}{C_L^{\kappa_\text{CMB} \kappa_\text{CMB}}}
&= 
2 \frac{\langle (\mathcal{Q}\left[ f_0, T_0 \right] + \mathcal{Q}\left[ T_0, f_0 \right])\ (\mathcal{Q}\left[ f_1, T_1 \right] + \mathcal{Q}\left[ T_1, f_1 \right])\rangle}{\langle \kappa_\text{CMB} \, \kappa_\text{CMB}\rangle} \\
&+ 
2 \frac{\langle (\mathcal{Q}\left[ f_1, T_0 \right] + \mathcal{Q}\left[ T_0, f_1 \right]).\ (\mathcal{Q}\left[ f_0, T_1 \right] + \mathcal{Q}\left[ T_1, f_0 \right])\rangle}{\langle \kappa_\text{CMB} \, \kappa_\text{CMB}\rangle} \\
\label{eq:secondary_calc}
\eal
\eeq
Here $\mathcal{Q}$ represents any quadratic lensing estimator, such as the QE, shear and magnification estimators.
This method was found to be much less noisy than the na\"ive approach of simply adding the lensed CMB and foregrounds and applying the lensing estimators to the sum. Furthermore, it does not involve any auto-spectrum, so no noise bias subtraction is needed.
This trick relies on the fact that the quadratic estimators are by construction unbiased when applied to $T^0$ and $T^1$.

For each foreground, we simulate 8060 flat square maps, of $30$ degrees and 1200 pixels on the side.
For the biases from lensed tSZ$\times$CIB, we ran only 6200 such simulations, due to limits in computing time.
This high resolution ensures that Fourier modes up to $2 \ell_\text{max, T}$ are correctly sampled in the maps, in order to avoid aliasing when nonlinear operations such as the first order lensing or the quadratic estimators are performed on the maps.

\section{Bias hardening against a lensed foreground}
\label{sec:bias_hardening}

In this section, we derive a modification to the standard CMB lensing quadratic estimator which nulls its response to foreground lensing.
We do this in the case of the standard quadratic estimator, but the same method can be generalized to the shear and magnification estimators.
As in the rest of the paper, we describe foreground lensing as a single foreground map being lensed by a single convergence map. This ignores the fact that each redshift slice of the foreground map is lensed by a slightly different lensing convergence field.

For a given lensing multipole $\vL$, the quantity
$\hat{\kappa}_{\vL, \vl} \equiv \frac{T_\vl T_{\vL-\vl}}{f^\text{CMB}_{\vl, \vL-\vl}}$
is an unbiased estimator of $\kappa_\vL^\text{CMB}$, 
with variance 
$\sigma_{\vL, \vl}^2 = \frac{2 C^\text{total}_\ell C^\text{total}_{|\vL-\vl|}}{f^\text{CMB\; 2}_{\vl, \vL-\vl}}$.
We denote $\sum_\vl \equiv \int \frac{d^2\vl}{(2\pi)^2}$,
and consider estimators of the form
$\hat{\kappa} = \sum_\vl w_\vl \hat{\kappa}_{\vl}$.

\subsection{Standard quadratic estimator}

The standard quadratic estimator of \cite{2002ApJ...574..566H} is simply the minimum variance unbiased linear combination of these estimators.
Indeed, we look for weights $w_{\vL, \vl}$ satisfying:
\beq
\left\{
\bal
&\sum_\vl w_{\vL, \vl} = 1 &&\text{ (unit response to CMB lensing)}\\
&\sum_\vl w_{\vL,\vl}^2 \sigma_{\vL, \vl}^2 \;\text{  is minimal} &&\text{(minimum variance)}\\
\eal
\right.
\eeq
This problem of minimization under constraints can be solved with Lagrange multipliers. We thus minimize the following quantity with respect to the lensing weights $w_{\vL, \vl}$ and the Lagrange multiplier $\alpha$:
\beq
\mathcal{L}=
\underbrace{\left( \sum_\vl w_{\vL,\vl}^2 \sigma_{\vL, \vl}^2 \right)}_\text{min. var.}
+\alpha\underbrace{\left( \sum_\vl w_{\vL, \vl} - 1 \right)}_{\text{unit response to }\kappa^\text{CMB}_\vL}
.
\eeq
This gives:
\beq
\hat{\kappa_\vL} \equiv 
\frac{\sum_\vl \hat{\kappa}_{\vL, \vl} / \sigma^2_{\vL, \vl}}
{\sum_\vl 1 / \sigma^2_{\vL, \vl}}
=
\frac{\int \frac{d^2\vl}{(2\pi)^2}\;
T_\vl T_{\vL-\vl}
\frac{f^\text{CMB}_{\vl, \vL-\vl}}{2 C^\text{total}_\ell C^\text{total}_{|\vL-\vl|}}}
{\int \frac{d^2\vl}{(2\pi)^2}\;
\frac{f^{\text{CMB }2}_{\vl, \vL-\vl}}{2 C^\text{total}_\ell C^\text{total}_{|\vL-\vl|}}},
\eeq
which is indeed the standard quadratic estimator of \cite{2002ApJ...574..566H}.

\subsection{Nulling the response to foreground lensing}

In the presence of a lensed foreground, the estimator $\hat{\kappa}_{\vL, \vl}$ acquires a bias $\mathcal{R}_{\vL, \vl}^f \, \kappa^f_\vL$, 
where 
$\mathcal{R}_{\vL, \vl}^f = \frac{f^\text{f}_{\vl, \vL-\vl}}{f^\text{CMB}_{\vl, \vL-\vl}}$.
Here we would like the combined estimator to have zero response to $\kappa^f_\vL$.
We thus look for lensing weights $w_{\vL, \vl}$ such that:
\beq
\left\{
\bal
&\sum_\vl w_{\vL, \vl} = 1 &&\text{ (unit response to CMB lensing)}\\
&\sum_\vl w_{\vL,\vl}^2 \sigma_{\vL, \vl}^2 \;\text{  is minimal} &&\text{(minimum variance)}\\
&\sum_\vl w_{\vL, \vl} \mathcal{R}_{\vL, \vl}^f  = 0 &&\text{ (zero response to foreground lensing)}\\
\eal
\right.
\eeq
We thus minimize the following quantity with respect to the lensing weights $w_{\vL, \vl}$ and the Lagrange multipliers $\alpha$ and $\beta$:
\beq
\mathcal{L}=
\underbrace{\left( \sum_\vl w_{\vL,\vl}^2 \sigma_{\vL, \vl}^2 \right)}_\text{min. var.}
+\alpha\underbrace{\left( \sum_\vl w_{\vL, \vl} - 1 \right)}_{\text{unit response to }\kappa^\text{CMB}_\vL}
+\beta\underbrace{\left( \sum_\vl w_{\vL, \vl} \mathcal{R}_{\vL, \vl}^f \right)}_{\text{zero response to } \kappa^\text{f}_\vL}
.
\eeq
The solution is:
\beq
w_{\vL, \vl} 
\equiv
\frac{1}{\left(S_0 S_2 - S_1^2\right)}
\frac{\left(S_2 - S_1 \mathcal{R}^f_{\vL, \vl}\right)}{\sigma^2_{\vL, \vl}},
\eeq
where the quantities $S_n$ are the following functions of $L$:
\beq
S_n \equiv
\int \frac{d^2\vl}{(2\pi)^2}\;
\left(\mathcal{R}_{\vL, \vl}^f\right)^n\,
\frac{f^\text{CMB\; 2}_{\vl, \vL-\vl}}{2 C^\text{total}_\ell C^\text{total}_{|\vL-\vl|}}.
\eeq
More explicitly, the lensed foreground bias hardened estimator is:
\beq
\hat{\kappa}_\vL
\equiv
\frac{1}{\left( S_0S_2 - S_1^2 \right)}
\int \frac{d^2\vl}{(2\pi)^2}\;
T_\vl T_{\vL-\vl}
\frac{f^\text{CMB}_{\vl, \vL-\vl}}{2 C^\text{total}_\ell C^\text{total}_{|\vL-\vl|}}\;
\left(S_2-S_1 \mathcal{R}_{\vL, \vl}^f\right),
\eeq
This estimator indeed has unit response to CMB lensing, zero response to the foreground lensing, and reduces to the standard quadratic estimator \cite{2002ApJ...574..566H} if the response $\mathcal{R}_{\vL, \vl}^f$ to foreground lensing is identically zero.

However, this lensed foreground bias hardened estimator has a larger noise power spectrum than the QE, given by:
\beq
N_\vL^{\hat{\kappa}}
=
\frac{1}{\left( S_0S_2 - S_1^2 \right)^2}
\int \frac{d^2\vl}{(2\pi)^2}\;
\frac{f^\text{CMB\; 2}_{\vl, \vL-\vl}}{2 C^\text{total}_\ell C^\text{total}_{|\vL-\vl|}}\;
\left(S_2-S_1 \mathcal{R}_{\vL, \vl}^f\right)^2.
\eeq
In principle, this approach could increase the noise by a large amount, in order to subtract the small bias due to foreground lensing.

\subsection{Minimizing the total variance from noise plus lensed foreground bias}

An alternative approach is thus not to require the response to foreground lensing to be exactly zero, but instead to minimize the total variance of the lensing estimator, including the additional variance due to the foreground lensing. This would make sure that we are not subtracting a $\sim 1\%$ foreground lensing bias at a large cost in signal-to-noise. 
In other words, we look for weights $w_{\vL, \vl}$ such that:
\beq
\left\{
\bal
&\sum_\vl w_{\vL, \vl} = 1 \text{  (unit response to CMB lensing)}\\
&\sum_\vl w_{\vL,\vl}^2 \sigma_{\vL, \vl}^2
+
2 \left( \sum_\vl w_{\vL,\vl} \mathcal{R}^f_{\vL, \vl} \right)
\left(  \sum_\vl w_{\vL,\vl'} \right)
C_L^{\kappa_f \kappa_\text{CMB}}
\;\text{  is minimal}\\
\eal
\right.
\eeq
The second line minimizes the sum of the noise variance plus the primary lensed foreground bias.
This ignores the secondary lensed foreground bias, which is important on the smaller lensing scales.
Again, we minimize the quantity:
\beq
\mathcal{L}=
\underbrace{\left( \sum_\vl w_{\vL,\vl}^2 \sigma_{\vL, \vl}^2 \right)}_\text{min. noise var.}
+
\underbrace{2 \left( \sum_\vl w_{\vL,\vl} \mathcal{R}^f_{\vL, \vl} \right)
\left(  \sum_\vl w_{\vL,\vl'} \right)
C_L^{\kappa_f \kappa_\text{CMB}}}
_\text{primary lensed foreground bias}
+\alpha\underbrace{\left( \sum_\vl w_{\vL, \vl} - 1 \right)}_{\text{unit response to }\kappa^\text{CMB}_\vL}.
\eeq
The solution is:
\beq
w_{\vL, \vl}
\equiv
\frac{1}{S_0 \sigma^2_{\vL, \vl}}
\left[
1
+ C_L^{\kappa_f \kappa_\text{CMB}}
\left( S_0 \mathcal{R}^f_{\vL, \vl} - S_1 \right)
\right] 
\eeq
The corresponding estimator is then:
\beq
\hat{\kappa}_\vL
\equiv
\frac{1}{S_0}
\int \frac{d^2\vl}{(2\pi)^2}\;
T_\vl T_{\vL-\vl}
\frac{f^\text{CMB}_{\vl, \vL-\vl}}{2 C^\text{total}_\ell C^\text{total}_{|\vL-\vl|}}\;
\left( S_0 \mathcal{R}^f_{\vL, \vl} - S_1 \right).
\eeq
Then estimator indeed has unit response to CMB lensing. It reduces to the standard quadratic estimator \cite{2002ApJ...574..566H} when $\mathcal{R}^f_{\vL, \vl} = 0$.
Its noise power spectrum is:
\beq
N_\vL^{\hat{\kappa}}
=
\frac{1}{S_0^2}
\int \frac{d^2\vl}{(2\pi)^2}\;
\frac{f^{\text{CMB }2}_{\vl, \vL-\vl}}{2 C^\text{total}_\ell C^\text{total}_{|\vL-\vl|}}\;
\left( S_0 \mathcal{R}^f_{\vL, \vl} - S_1 \right)^2.
\eeq

\end{document}